

%
%
%
%

\input harvmac.tex
\input epsf
\def\LG{Lan\-dau-Ginz\-burg\ }
\def\gkl{{\cal G}_{k,\ell}}

\def\half{{1 \over 2}}

\def\cmp#1{{\it Commun.\ Math. \ Phys.} \ {\bf #1,\/}}
\def\nup#1{{\it Nucl.\ Phys.} \ {\bf B#1,\/}}
\def\plt#1{{\it Phys.\ Lett.}\ {\bf #1B,\/}}
\def\ijmp#1{{\it Int. \ J.\ Mod. \ Phys.}\ {\bf A#1,\/}}

\def\Gminus{G_{-{1 \over 2}}^-}
\def\Gplus{G_{-{1 \over 2}}^+}
\def\Gkone{{\cal G}_{k,1}}
\def\inbar{\vrule height1.5ex width.4pt depth0pt}
\def\IC{\relax\,\hbox{$\inbar\kern-.3em{\rm C}$}}
\def\IP{\relax{\rm I\kern-.18em P}}
\def\IR{\relax{\rm I\kern-.18em R}}
\font\sanse=cmss12
\def\ZZ{\relax{\hbox{\sanse Z\kern-.42em Z}}}
\def\binv{{ 1 \over \beta}}

\font\ninerm=cmr9

%
\def\Titletwo#1#2#3#4{\nopagenumbers\abstractfont\hsize=\hstitle
\rightline{#1}\rightline{#2}
\vskip .7in\centerline{\titlefont #3}
\vskip .1in \centerline{\abstractfont {\titlefont #4}}
\abstractfont\vskip .5in\pageno=0}

\def\Date#1{\leftline{#1}\tenpoint\supereject\global\hsize=\hsbody%
\footline={\hss\tenrm\folio\hss}}
%

\Titletwo{}{}
{Lattice Analogues of $N=2$ Superconformal Models}{via Quantum Group
Truncation$^*$
{\abstractfont
\footnote{}{$^*$ Work supported in part by funds provided by the DOE
under grant No. DE-FG03-84ER40168.}} } {}
\centerline{Z. Maassarani, D. Nemeschansky \ and \ N.P. Warner
\footnote{$^\dagger$}{ Alfred P. Sloan foundation fellow.}}
\bigskip \centerline{Physics Department}
\centerline{University of Southern California}
\centerline{University Park}
\centerline{Los Angeles, CA 90089-0484.}
\vskip 1.0cm
We obtain lattice models whose continuum limits correspond to
$N=2$ superconformal coset models. This is done by taking the
well known vertex model whose continuum limit is the
$G \times G/G$ conformal field theory, and twisting the
transfer matrix and modifying the quantum group truncation.
 We find that the natural order parameters
of the new models are precisely the chiral primary fields.
The integrable perturbations of the conformal field
theory limit also have natural counterparts in the lattice
formulation,
and these can be incorporated into an affine quantum group structure.
The topological, twisted $N=2$ superconformal models also have
lattice analogues, and these emerge as an intermediate part of our
analysis.

\vskip .1in
\vfill\leftline{USC-92/007 }
\Date{April 1992}
%


\newsec{Introduction}

In this paper we describe some  exactly solvable
lattice models whose continuum limit yields the $N=2$
superconformal coset models based on hermitian symmetric spaces
\ref\KS{Y.~Kazama and H.~Suzuki, \nup{234} (1989) 73.}.
Our approach is, to some degree, a natural extension of the
lattice analogues of the $N=2$ supersymmetric minimal
series via fusions of the six vertex model
\ref\HS{H.~Saleur, Polymers and Percolation in two Dimensions and
Twisted $N=2$ supersymmetry Yale preprint YCTP-P38-91, (1991);
H.~Saleur,  Geometrical Lattice Models for $N=2$ Supersymmetric
Theories in two Dimensions Yale preprint YCTP-P39-91 (1991). }.
We do not use a manifestly supersymmetric formulation:
instead the lattice formulation employed here parallels the free
bosonic
and para-toda formulation of the $N=2$ superconformal models
\ref\FLMW{P.~Fendley, W.~Lerche, S.~Mathur and N.P.~Warner
\nup{348} (1991) 66.}
\ref\EHY{T.~Eguchi, S.~Hosono and S.-K.~Yang, {\it Commun. Math.
Phys.} 140 (1991) 159.}
\ref\JAP{T.~Eguchi, T.~Kawai, S.~Mizoguchi and S.K.~Yang,
Tokyo preprint KEK-TH-303
(1991); Character Formulas for $N=2$ Superconformal Theories. }
\ref\NW{D.~Nemeschansky and N.P.~Warner; USC preprint USC-91/031
(1991);
Topological Matter, Integrable Models ans Fusion Rings,
to appear in {\it Nucl. Phys. B } }.
The starting point of our construction is the well known vertex
model,
whose continuum limit is a $G_k \times G_1 /G_{k+1}$
conformal coset model.
By passing to a topological model and ``untwisting'' one can get  a
class of $N=2$ superconformal coset models \EHY\NW .
We translate this prescription into the construction of lattice model
analogues
of these $N=2$ superconformal theories.
It is unclear whether the lattice model itself has any form of
supersymmetry
even though the continuum limit does.
We do, however, suspect that there is some hidden supersymmetry since
the
lattice models we describe can be ``twisted'' back into topological
lattice models that are the direct analogues of the topological,
twisted
$N=2$ superconformal theories of \ref\twist{E. Witten,
{\it Commun. Math. Phys.}  {\bf 117} (1988) 353; T. Eguchi,
S.-K. Yang, {\it Mod. Phys. Lett.}  {\bf A4} (1990) 1653.}.
We suspect that it would not be so simple to obtain such topological
lattice models unless there were some kind of hidden
supersymmetry generators that provide the
truncation to the topological lattice models.

Once the lattice analogue of an $N=2$ superconformal coset is
obtained,
one finds that it comes equipped with a set of natural operators.
These operators appear as an affine extension of the underlying
quantum group structure of the vertex model, and we identify these
operators
as lattice analogues of the most relevant supersymmetric
perturbations
of the $N=2$ superconformal model.
In the continuum limit such perturbations lead to a
massive integrable field theory \ref\FMVW{P.\ Fendley, S.\ Mathur,
C.\ Vafa and N.P. \ Warner, \plt{243} (1990) 257.}\FLMW.
It also turns out to be straightforward to
identify the natural lattice order
parameters, and we find that the corresponding operators renormalize
to
Landau-Ginzburg fields of the $N=2$ superconformal
coset model in the continuum limit.

In this paper we will concentrate on the so-called $SLOHSS$ models
\foot{This miserable acronym stands for Simply Laced, Level
One, Hermitian Symmetric Space models.},
but our results can readily be generalized via fusion procedure to
arbitrary $N=2$ supersymmetric models based upon hermitean symmetric
spaces.
We begin in section 2 by reviewing relationships between
coset models $\Gkone \equiv G_k \times G_1 /G_{k+1} $
and the $SLOHSS$ models.  We then exploit this in section 3  to
construct the new lattice models in both the vertex and
restricted height formulations.  The lattice analogues of the
grassmannian models are described in some detail  and the Boltzmann
weights are computed in the restricted height formulation.
In section 4 we discuss various lattice operators, and in section 5
we discuss some of the questions that are raised by this work.

\newsec{Coulomb gas formulation of $SLOHSS$ models}

We consider the $N=2$ superconformal coset models of the form
\eqn\cosetmodel{{ G_1 \times SO(dim(G/H)) \over H } \ ,}
where $G$ is simply laced and of level one,
and $G/H$ is a hermitean symmetric space.
The free field formulation has been discussed extensively elsewhere
\FLMW\EHY\JAP\NW , but to summarize, and fix
notation, we consider $\ell$ free bosons, $\phi$,
(where $\ell$ is the rank of $G$).
The stress tensor, $T(z)$, has the form
\eqn\stresstensor{T(z) \ = \ - \half (\del \phi(z))^2 + i
\big ( \beta -\binv \big)
\ \rho_H \cdot \del^2 \phi(z) \ \ , }
where
\eqn\defofbeta{ \beta  \ = \ \sqrt{ g \over g+1}\ ,}
$\rho_H$ is the Weyl vector of the subgroup $H$ of $G$, and $g$ is
the
dual Coxeter number of $G$.
The central charge of the model is $c= {3M \over g+1 }$, where $M$ is
the
complex dimension of $G/H$.
The screening currents are
\eqn\screeners{ e^{i \beta \alpha_j \cdot \phi(z)} \ \ \ {\rm and}
\ \ \ \ \ e^{-i \binv \alpha_j \cdot \phi(z) }\ , \ \ \ \ \ j=1,
\ldots , \ell -1 \ , }
where the $\alpha_j$ are the simple roots of $H$.
The two supercurrents, $G^+(z)$ and $G^-(z)$, can
be represented by vertex operators,
\eqn\susycharges{G^+(z) \ = \ e^{-i \binv \gamma \cdot \phi(z)}
\ \ \ \ \ \ G^-(z) \ = \ e^{i \binv \psi \cdot \phi(z)} \ , }
where $\gamma$ is the simple root of $G$ that extends the simple root
system of $H$  to a simple root system for $G$, and $\psi$ is the
highest
root of $G$.
The $U(1)$ current of the $N=2$ superconformal algebra is given by,
\eqn\uonecurrent{ J(z) \ = \ 2 i \big( \beta - \binv \big)
( \rho_G-\rho_H )\cdot \del \phi \ , }
where $\rho_G$ is the Weyl vector of $G$.

The chiral primary fields are easily identified \ref\LVW{W.~Lerche,
C.~Vafa and N.P~Warner, \nup{324} (1989)427. } \ref\Gep{ D.~Gepner
\plt{222} (1989) 207; ``A Comment on the Chiral Algebra of Quotient
Superconformal Field Theory'', Princeton preprint PUPT 1130 (1989);
{\it Commun. Math. Phys.} {\bf 142} (1991) 433.}
\ref\WLNW{W.~Lerche and N.P.~ Warner, \nup{358} (1991) 571.}.
They can all be represented by vertex operators,
\eqn\chiralprimary{ e^{-i(\beta-\binv) \Lambda \cdot \phi(z)
} \ \ \ {\rm with} \ \ \ \Lambda \ = \ w(\rho_G) ~-~ \rho_G \ ,}
where $w$ an element of the Weyl group, $W(G)$, of $G$ is chosen so
that $w(\rho_G)$ is a dominant weight of  $H$.
The vertex operator \chiralprimary\ has $N=2$, $U(1)$ charge:
$$Q ~=~ {1 \over {g(g+1)}} ~ 2(\rho_G - \rho_H)\cdot
[\rho_G - w(\rho_G)] \ .$$

The toplogically twisted $N=2$ superconformal model is obtained
from the $N=2$ superconformal model by taking \twist:
\eqn\topT{T_{top} (z) \ = \ T(z) ~+~ \half \partial J(z)
{}~=~ - \half (\del \phi)^2 + i  \big ( \beta -\binv \big)
\ \rho_G \cdot \del^2 \phi \ \ , }
and using $G^+(z)$ (which now has conformal weight equal to $1$)
as a screening current.  The physical states of the topological model
precisely the chiral primary fields \chiralprimary.

The partition functions of the original (untwisted) $N=2$
superconformal field theory can easily be computed
 using the Coulomb gas
formulation \ref\GFel{G.~Felder, \nup{317} (1989) 215; Erratum
\nup{324} (1989) 548.}
\ref\BMP{P.~Bouwknegt, J.~McCarthy and K. ~Pilch, \nup{352} (1991)
139; {\it Prog. Theor. Phys. Suppl.} {\bf 102} (1990) 67.}.
One can also obtain them from a direct computation of
the branching functions \ref\KW{V.G.~Kac and M.~Wakimoto,
{\it  Proc. Natl.  Acad. Sci. USA} {\bf 85} (1988) 4956;
{\it Acta Applicandae Math.} {\bf 21} (1990) 3;
P.~Christe and F. ~Ravanini, {\it Int. J. Math. Phys.}
{\bf A4} (1989) 897.}.
Since the $SLOHSS$ models do not have any fixed points
under the field identifications generated by spectral flow, the
branching
functions can be identified with the characters of the model
\KS\ref\MS{G.~Moore and  N.~Seiberg, \plt{220} (1989) 422.}
\ref\DGep{D.~Gepner, \plt{222} (1989) 207.}\LVW
\ref\SY{B.~Schellekens and S.~Yankielowicz, \nup{334} (1990) 67.}.
Following \KS, we introduce $M$ complex fermions,
$\lambda^{\bar\alpha}(z)$ and $\lambda^{-\bar\alpha}(z)$, to describe
$SO_1(dim(G/H))$.
The labels, $ \bar\alpha$, are the positive roots of $G$ that are not
positive roots of $H$.
In order to calculate the branching functions of the model
we need to analyze the conformal
embedding of $H$ into $SO(dim(G/H))$.
The Cartan subalgebra of $H$ is realized via
\eqn\cartansu{h^i(z) \ = \ \sum_{\bar\alpha}\ \bar\alpha^i ~
\lambda^{-\bar\alpha}(z) \lambda^{\bar\alpha}(z) \ . }
The fermionic partition function, with an insertion of $e^{-2\pi i
\nu_i
 h_0^i}$, are
\eqn\fermichar{\eqalign{ \chi_R^{\pm}(\tau;\nu) \ & = \ q^{M/12}
\prod_{\bar\alpha} \Big[
(e^{-i\pi \bar\alpha\cdot\nu}
\pm e^{i\pi \bar\alpha\cdot\nu})
\prod_{n=1}^{\infty}(1\pm q^n e^{-2\pi i \bar\alpha
\cdot\nu})(1\pm q^n e^{2\pi i \bar\alpha\cdot\nu } )\Big] \cr
\chi^\pm_{N.S.} (\tau;\nu) \ & = \ q^{-M/24} \prod_{\bar\alpha}
\prod_{n=1}^{\infty}(1\pm q^{n-\half} e^{-2\pi i
\bar\alpha\cdot\nu})(1\pm q^{n-\half}e^{2\pi i \bar\alpha\cdot\nu})
\ . \cr}}
These characters are interrelated  by spectral flow:
\eqn\specflow{\eqalign{\chi_{N.S.}^{\pm}(\tau; \nu+{1 \over g
}(\rho_G
-\rho_H))\ & = \ \chi_{N.S.}^\mp (\tau;\nu) \cr
\chi_R^{\pm}(\tau;\nu +{1 \over g }(\rho_G-\rho_H)) \ & =\
(-i)^M\chi_R^{\mp}(\tau;\nu) \cr
\chi_{N.S.}^{\pm}(\tau;\nu+{1\over g}(\rho_G-\rho_H)\tau) \ & =\
(\pm1)^Mq^{-M/8} e^{-2\pi i (\rho_G-\rho_H)\cdot \nu }
\chi^{\pm}_R(\tau;\nu) \cr
\chi^{\pm}_R(\tau; \nu +{1\over g} (\rho_G-\rho_H)\tau ) \ & =
\ q^{-M/8} e^{-2 \pi i (\rho_G-\rho_H)\cdot \nu} \chi_{N.S.}^{\pm}
(\tau;\nu) \ . \cr}}
More significantly, if $\Delta(G)$ is the Weyl-Kac denominator of
$G$:
\eqn\Weyldom{\eqalign{\Delta(G) \ = \ & \prod_{\alpha \in\Delta^+(G)}
\Big[ (e^{-i \pi \alpha \cdot \nu} -e^{i\pi \alpha\cdot \nu})
\cr & \times \prod_{n=1}^\infty (1-q^n)^\ell (1-q^n
e^{-2\pi i \alpha \cdot \nu})
 (1- q^n e^{2 \pi i \alpha \cdot \nu}) \Big]  \ ,\cr } }
then one can write:
\eqn\chir{ \chi_R^-(\tau;\nu)  \ = \ {\Delta(G) \over \Delta(H)} \ .
}
Using the Weyl-Kac dominator formula for $G$ one immediately obtains:
\eqn\chirweyl{\chi_R^{\pm}(\tau;\nu) \ = \ q^{M/12} {1 \over
\Delta(H)}
\sum_{w \in W(G) } \sum_{\alpha \in M(G)} \epsilon^{\pm}(w,\alpha)
q^{{1 \over 2g}
[w(\rho_G)+g\alpha]^2 -{1 \over 2g}\rho_G^2} e^{-2 \pi i \nu\cdot
[w(\rho_G)+g\alpha]} \ ,}
where $M(G)$ is the root lattice of $G$, and
\eqn\defepsilon{\eqalign{ \epsilon^-(w,\alpha) \ & = \ \epsilon (w)
\cr
\epsilon^+(w,\alpha) \ & = \ \epsilon (w) e^{-{2 \pi i \over g}
(\rho_G-\rho_H)\cdot [w(\rho_G)-\rho_G+g\alpha]} \ .
\cr }}
Using \specflow\ one then has
\eqn\chinsweyl{\eqalign{\chi_{N.S.}^{\pm}(\tau;\nu) \ = \  q^{M/12}
{1
\over \Delta(H)}
\sum_{w \in W(G) } \sum_{\alpha \in M(G)}
& \epsilon^{\pm}(w,\alpha)q^{{1 \over 2g}
[w(\rho_G)+g\alpha-(\rho_G-\rho_H)]^2 -{1 \over 2g}\rho_G^2} \cr
& \times e^{-2 \pi
i \nu\cdot [w(\rho_G)+g\alpha-(\rho_G-\rho_H)]} \ .\cr }}

{}From this one can reduce the highest weight representations of
affine
$SO_1(dim(G/H))$ into finitely many such representations of
$H_{g-h}$.
(The number $h$ is the dual Coxeter number of $H$, and should be
thought
of as a vector if $H$ is semi-simple.  The dual Coxeter number of the
$U(1)$ factor is defined to be zero.)
In the Ramond sector we have
\eqn\ramon{\chi_R^{\pm}(\tau;\nu) \ = \ \sum_{w \in {W(G) \over
W(H)}}
\sum_{\alpha \in {M(G) \over M(H)}}
\chi_{\lambda(\alpha,w)}^{H_{g-h}}(\tau ;\nu ) \ , }
where
\eqn\deflambda{\lambda(\alpha,w) \ = \ w(\rho_G) -\rho_H +g \alpha }
and $\chi^{H_{g-h}}_\lambda$ is the character  of $H$ at level $g-h$
with highest weight $\lambda$.
The Weyl element, $w \in {W(G) \over W(H)}$, is chosen so that
$\lambda(\alpha, w) $ is a highest weight of $H$.

Now introduce the level one characters of $G$
\eqn\levelone{ \chi^G_{\Lambda}(\tau;\nu) \ = \ {1 \over
{\eta(\tau)^\ell} } \sum_{\alpha \in M(G) } q^{{1 \over
2}(\Lambda + \alpha)^2}
 e^{-2\pi i \nu \cdot (\Lambda + \alpha ) } \ , }
and multiply this by the expressions \chirweyl\ and \chinsweyl\
for $\chi_R^\pm$ or $\chi_{N.S.}^\pm$.
By rearranging the sum it is elementary to factor out the branching
functions.
At this point it is also valuable to keep track of the $N=2$ , $U(1)$
current.
One  then obtains the branching functions for the $N=2$ Hilbert space
with an insertion of $e^{-2\pi i \zeta J_0 }$ into the trace:
\eqn\branching{\eqalign{b_{\Lambda, R_\pm}^{\lambda_-}\ = \  {1\over
\eta(\tau)^\ell}\sum_{w\in W(G)} & \sum_{ \alpha \in M(G)}
 \epsilon^\pm(w,\alpha )~q^{\half \big [{1 \over \beta}  w(\rho_G)
- \beta (\lambda_-+\rho_H)+\sqrt{g(g+1)}\alpha\big ]^2} \cr
&\times e^{-{4\pi i \over \sqrt{g(g+1)}}\zeta (\rho_G-\rho_H)\cdot
\big [{1 \over \beta}  w(\rho_G) - \beta (\lambda_- +
\rho_H)+\sqrt{g(g+1)}\alpha \big]} \cr }}
\eqn\branchingns{\eqalign{b_{\Lambda, N.S._\pm}^{\lambda_-}\ = \
{1\over
\eta(\tau)^\ell }\sum_{w\in W(G)} & \sum_{ \alpha \in M(G)}
\epsilon^\pm(w,\alpha )~q^{\half \big[{1 \over \beta}
( w(\rho_G)-\rho_G+\rho_H)- \beta (\lambda_-+\rho_H)+
\sqrt{g(g+1)}\alpha \big ]^2} \cr
& \times e^{-{4\pi i \over \sqrt{g(g+1)}}\zeta
(\rho_G-\rho_H)\cdot \big [{1 \over \beta} ( w(\rho_G)-\rho_G+\rho_H)
- \beta  (\lambda_-+\rho_H)+\sqrt{g(g+1)}\alpha \big ]}\  \cr }}
where $\beta $ is given by \defofbeta.
Taking the usual $A$-type modular invariant we obtain the modified
gaussian partition function
\eqn\gaussian{\eqalign{Z \ & = \ {1 \over 2 |Z(G)|} \sum_{{\Lambda,
\lambda_- \atop u\in \{R_{\pm}, N.S._{\pm}\} }}
|b_{\Lambda, u}^{\lambda_-}|^2
\cr & = \ {1 \over 2 |W(H)||Z(G)| }~{ 1 \over | \eta(\tau) |^{2\ell}
}
\sum_{w \in W(G)} \sum_{v
\in {\binv M(G)^*\over \Gamma} } \sum_{u\in {\beta M(G)^* \over
\Gamma} }
\sum_{v_1, v_2 \in \Gamma} \sum_{{\xi =0,1 \atop  \eta = 0,1}}
\epsilon (w) \cr &\qquad \qquad
 \times q^{\half (v_L+\eta s)^2}\bar q^{\half(v_R+\eta s)^2}
e^{-4\pi i s\cdot (\zeta_L  v_L - \zeta_R  v_R)}
 e^{-2\pi i \xi s \cdot ( v_L- v_R)}\ , \cr }}
where
\eqn\defgamma{\Gamma \ = \  \sqrt {g(g+1) }M(G) \ }
\eqn\vlvr{ v_L \ = \  v + u + v_1 \ ; \ \ \ \ v_R \ = \ w(v)+u+v_2 \
}
and
\eqn\defs{s \ = \  {1 \over
\sqrt{g(g+1)}}( \rho_G-\rho_H) \ .}
The partition function \gaussian\ represents $Tr [q^{L_0- c /24} \bar
q^{\bar L_0 - c/24} e^{-2\pi i (\zeta_L J_0 -\zeta_R \bar J_0 )}]$
taken over the entire Hilbert space, and the factor of $|Z(G)|^{-1}$
is a division by the order of the center of $G$ and factors out
the field identifications induced by the spectral flow by the center
 \MS\DGep\LVW.  In \gaussian, the
sum over  $\eta = 0$ and $\eta = 1$ corresponds to the
sum over the Neveu-Schwarz and Ramond sectors respectively.  The
sum over  $\xi = 0$ and $\xi = 1$ corresponds, respectively,
to the insertion or absence of $(-1)^F$ in the trace.
It is, perhaps, also, amusing to note that for fixed $w \in W(G)$
the sum over $(v_L;v_R)$ in \gaussian\ is a sum over a
Lorentzian self-dual lattice and thus \gaussian\ is modular
invariant even before one sums over $W(G)$.
The sum over $\eta$ and $\xi$ is a simple example of the shifted
lattice
construction that maps one modular invariant to another
\ref\LSW{W.~Lerche, A.N.~Schellekens and N.P.~Warner, {\it Phys.
Rep.}
{\bf 177} (1989) 1.}.

It is now highly instructive to compare the $N=2$ superconformal
model
\cosetmodel\ to the coset model $\Gkone \equiv G_k \times
G_1/G_{k+1}$. We first note that the branching functions and
partition
functions can be computed in the same manner as for the $N=2$
supersymmetric models\KW.  The diagonal modular
invariant partition function for the $\Gkone$ model is given by
a very similar formula to \gaussian:
\eqn\gaussiang{Z \ =
\   {1 \over  |W(G)||Z(G)| } { 1 \over | \eta(\tau) |^{2\ell}}
\sum_{ w \in W(G)} \sum_{v
\in {\binv M(G)^*\over \Gamma }  } \sum_{u \in {\beta M(G)^* \over
\Gamma}}
\sum_{v_1,v_2 \in \Gamma }  \epsilon(w) \
  q^{\half (v_L)^2}\bar q^{\half(v_R)^2}}
where one now has
\eqn\newbeta{ \beta  \ = \ \sqrt{ {k + g} \over {k + g + 1}}\ ,}
\eqn\newgamma{\Gamma \ = \  \sqrt {(k+g)(k+g+1) }M(G) \ , }
but $v_L$ and $v_R$ are still given in terms of $u$, $v$, $v_1$
and $v_2$ by \vlvr.

This form of the partition function can also be obtained from the
well known free bosonic formulation of the $\Gkone$ coset models
\ref\FLy{V.A.~Fateev and S.L.~Lukyanov, \ijmp{3} (1988) 507;
``Additional symmetry and exactly-soluble models in two
dimensional conformal field theory'', Landau Institute preprint
 (1988); T.~Hollowood and P.~Mansfield, \plt {226} (1989) 73;
P.~Bouwknegt, J.~McCarthy and K.~Pilch, \nup{352} (1991) 139.}.
In this formulation, the
energy momentum tensor is
\eqn\genmom{T_G (z) \ = \  - \half (\del \phi)^2 + i
\big ( \beta -\binv \big) \ \rho_G \cdot \del^2 \phi \ \ , }
where $\beta$ is given by \newbeta.  The screening currents of
the $\Gkone$ model are $e^{i \beta \alpha_j \cdot \phi}$ and
$ e^{-i \binv \alpha_j \cdot \phi }$ for $j=1, \ldots , \ell$,
where $\alpha_j$ are the simple roots of $G$. (In the notation
following equation \susycharges\ we are taking $\alpha_\ell
\equiv \gamma$.)  The primary fields can be represented by
\eqn\gprims{\Phi_{\Lambda_1, \Lambda_2} ~=~ e^{(-i
(\beta \Lambda_2 ~-~ \binv \Lambda_1 ) \cdot  \phi(z) )} \ , }
where $\Lambda_1$ is an affine highest weight of $G_k$ and
$\Lambda_2$ is an affine highest weight of $G_{k+1}$.  The labels
$\Lambda_1$ and $\Lambda_2$ also directly correspond,
in the obvious way, to the branching functions of $\Gkone$.
(The label of $G_1$ in the numerator of the coset is equal
to $\Lambda_2 - \Lambda_1$ modulo the root lattice of $G$).

One should note that the topologically twisted $N=2$ superconformal
model is almost the same as the a $\Gkone$ model with $k=0$
\ref\WL{W.~Lerche, \plt{252} (1990) 349.}\EHY, but with an important
modification - the ${\cal G}_{0,1}$ model is a very particular,
supersymmetry preserving perturbation of the twisted $N=2$ model \NW.
This perturbation is generated by the use of the screening current
$e^{-i \binv \gamma \cdot \phi(z)}$.  We will remark further upon
this perturbation in  section 4.

To make contact with the lattice model one can Poisson re-sum
\gaussian\ and \gaussiang\ over the lattice
$u+v_1 \in \beta M^*(G)$ .
For the $N=2$ superconformal  partition function one finds
\eqn\poissonr{\eqalign{Z  =   { {A} \over
{(2Im (\tau))^{{\ell \over 2}}\ |\eta (\tau) |^{2\ell} }}
\sum_{x \in \binv M(G)}& \sum_{v \in {\binv M^*(G) \over \Gamma}}
\sum_{v_0 \in \Gamma}\sum_{\xi, \eta =0,1}
 e^{-{\pi \over 2 Im (\tau)} |x- \tau (v-w(v) +v_0) -
2(\zeta_L-\zeta_R)s|^2} \cr &
\times e^{i \pi x\cdot (v+w(v) -v_0+2\eta s)}
e^{-2\pi i(\zeta_L+\zeta_R+\xi)s\cdot (v-w(v) +v_0)} \cr}}
where $A = (2 |W(H)| |Z(G)|^\half \beta^\ell )^{-1}$ is an
irrelevant normalization constant.
The result for $\Gkone \equiv G_k \times G_1/G_{k+1}$ is identical
(up to another normalization constant ), but with
$\xi=\eta=\zeta_L=\zeta_R=0$ and $\beta$ and $\Gamma$ given by
\newbeta\
and \newgamma.

It has been convincingly argued \ref\FSZO{P.~Di~Francesco, H.~Saleur
and J.-B.~Zuber, J. Stat. Phys {\bf 34} (1984) 731; J. Stat. Phys.
{\bf 49} (1987) 57; \nup{285} (1987) 454.} \ref\FSZ{P.~Di ~Francesco,
H.~Saleur and J.-B. Zuber, \nup{300} (1988) 393;
J.-B.~ Zuber, ``Conformal Field Theories, Coulomb Gas Picture
and Integrable models'', in {\it Fields, Strings and Critical
Phenomena},
editors E.~Br\'ezin and J.~Zinn-Justin, Les Houches 1988 Session
XLIX.}
\ref\IK{I.~Kostov, \nup{300} (1988) 559.}
that (at least for $G=SU(n)$) the Poisson re-summed form of the
partition function \gaussiang\ corresponds to that obtained from the
$IRF$
model based on the representations of $G$.
(This will be described more fully in the next section).
The key observation that we wish to make here is that the
SLOHSS model is a very simple  modification of the $G\times G /G$
model.
To go from  $\Gkone$ to \cosetmodel, one sets $k=0$ and then untwists
the resulting topological energy-momentum tensor \topT.
In the continuum theory this
means subtracting the total derivative $\half \partial J(z)$ from
\genmom\ (with $k=0$).  In the next section we will see that
this amounts to choosing a particular value of $q$ and modifying
a boundary term in the transfer matrix of the lattice model.  The
only
other changes that are necessary in order to get the model
\cosetmodel\ are some appropriate phase insertions, and these
can be read off from \poissonr\ and its counterpart in the
$\Gkone$ model.  Specifically, in the partition function \poissonr,
each term in the soliton sum is given an additional phase that
depends upon the $SO(dim(G/H))$, or fermionic, sector of \cosetmodel.
This phase is:
\eqn\newphase{e^{2\pi i [\eta x\cdot s \ -\ \xi y\cdot s]},
 \quad y= v -w(v) + v_0\ ,}
where $s$ is defined in \defs\ and the significance of $\eta$ and
$\xi$ is described below \defs.
In terms of the lattice model, the momentum vectors $x$ and $y$ in
\poissonr\ and \newphase\ represent
the change in the height, $\Delta \phi$, as one
periodically identifies the lattice.  The extra phases, \newphase,
must therefore be properly incorporated
in order to obtain the desired  lattice model from the lattice
model corresponding to $\Gkone$.

\newsec{Lattice models}

The relationship between the conformal coset models
\eqn\gklcoset{\gkl \ = \ { G_k \times G_{\ell} \over G_{k+\ell}}\ , }
and lattice
models is fairly well established
 \ref\latt{A.A.~Belavin, \nup{188} (1981) 189;
O.~Babelon, H.~de~Vega and C.~Viallet \nup{190} (1981) 542;
\nup{200} (1982) 266;  P.P.~Kulish, N.Yu.~Reshetikhin and E.K.~
Sklyanin, {\it Lett. Math. Phys.} {\bf 5} (1981) 393;
M.~Jimbo, T.~Miwa and M.~Okada, {\it Lett. Math. Phys}
{\bf 14} (1987) 123; {\it Commun. Math. Phys.} {\bf 116}
(1988) 507; M.~Jimbo, A.~Kuniba,  T.~Miwa
and M.~Okada, {\it Commun. Math. Phys.} {\bf 119} (1988) 543. }
\ref\PA{V.~Pasquier, \nup{295} (1988) 491. }
\ref\FZU{P.~Di~Francesco and J.B.~Zuber, \nup{338} (1990) 602.}.
The conformal model only appears in the continuum limit
and at the critical point of the  lattice model.
For simplicity we will only consider the critical transfer matrices
here.
For $\ell =1$, the vertex description of the lattice model is
obtained by
building the  transfer matrix from the $\check R$-matrix for the
fundamental representation of $G$, and then performing a
quantum group truncation \foot{We will be considering the vertex
model
transfer matrix with free boundary conditions.
To discuss the vertex model
transfer matrices with periodic boundary conditions one
has to do some unpleasant technical modifications. See, for example,
\ref\PS{V.~Pasquier and  H.~Saleur, \nup{330} (1990) 523.} . }
\PA\PS\ref\SaZu{ H.~Saleur and J.-B. Zuber,
``Integrable Lattice Models and Quantum Groups'', in the proceedings
of the 1990 Trieste Spring School on String Theory and Quantum
Gravity.}.
One considers evolution from left to right on the
usual $45^0$ lattice (see figure 1).
A constant time-slice is a vertical zig-zag that runs from the top
to the bottom of the lattice.
Suppose this zig-zag has $2L$ edges.
To each edge one associates a copy of
the fundamental representation, $V$, of $G$ and the Hilbert space of
the
time slice is ${\cal V} = V^{\otimes 2L}$.

\epsfxsize = 3.5in
\vbox{\vskip -.1in\hbox{\centerline{\epsffile{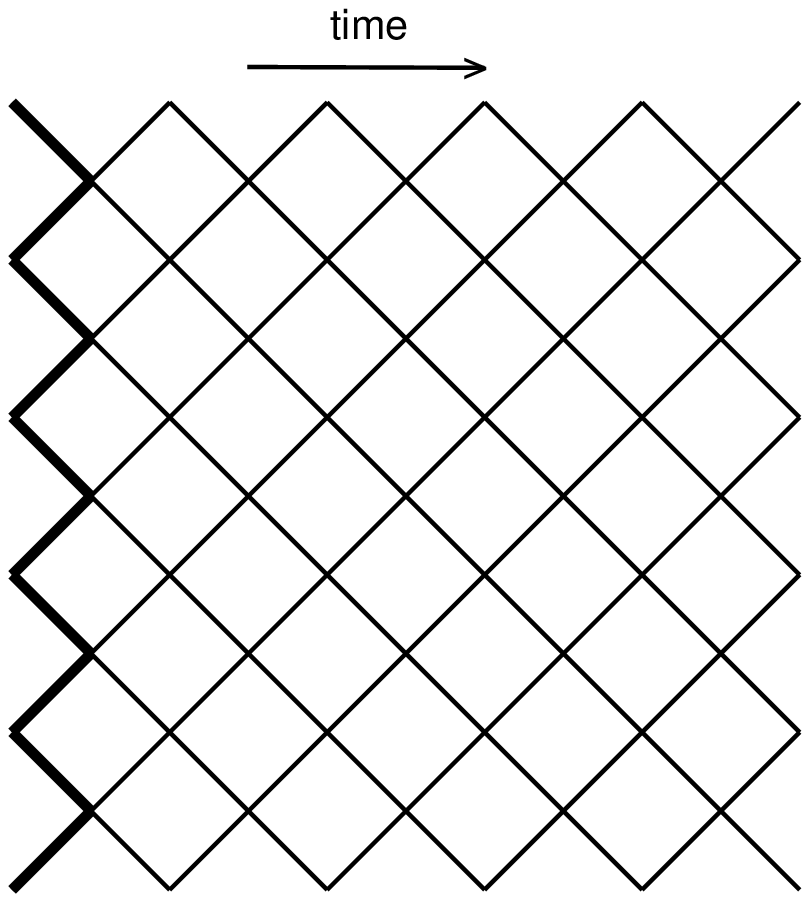}}}
\vskip -.5in
{\leftskip .5in \rightskip .5in \noindent \ninerm \baselineskip=10pt
Figure 1. A section of the lattice upon which the vertex
model is defined.  The bold zig-zag is the initial time slice.
\smallskip}} \bigskip \bigskip

The transfer matrix is given by
\eqn\tranmatrix{ {\cal T }(u,q) \ = \ \Bigl [ \prod_{p=1}^L
X_{2p-1}(u,q) \Bigr ] \Bigl [ \prod_{p=1}^{L-1}
X_{2p}(u,q) \Bigr ]  \ , }
where
\eqn\defxi{ X_p(u,q) \ = \ { 1 \over 2i }~ \check R(u,q) \ , }
and this matrix acts on the tensor product
of the $p^{th}$ and $(p+1)^{th}$ copy of $V$. The matrix,
$\check R(u,q)$, is the $\check R$-matrix for quantum $G$,
and maps $V\otimes V \rightarrow V \otimes V$.
For $SU(n)$, $V$ is  the $n$-dimensional representation,
and one has \ref\JMBO{M.~Jimbo,
{\it Commun. Math. Phys.} {\bf 102} (1986) 537.}
\eqn\chekrmatic{\eqalign{\check R(u,q) \ = \ & (xq-x^{-1}q^{-1})
\sum_{\alpha=1}^n E_{\alpha \alpha} \otimes E_{\alpha \alpha} + (x-
x^{-1}
 ) \sum_{ { \alpha \neq\beta\atop \alpha, \beta = 1 }}^n E_{\alpha
\beta} \otimes E_{\beta \alpha} \cr
& +  (q-q^{-1} )\Bigl [ x \sum_{ \alpha > \beta}  ~+~ x^{-1}
\sum_{\alpha < \beta} \Bigr ] E_{\alpha \alpha }
\otimes E_{\beta\beta } \ , \cr }}
where $x=e^{iu}$, $E_{\alpha\beta}$ is an $n\times n$ matrix whose
entries
$(E_{\alpha \beta})_{ij}$ are equal to $\delta_{\alpha i }
\delta_{\beta j
}$.
Literally, by construction, the transfer matrix \tranmatrix\ commutes
with
the action of $U_q(G)$ on $\cal V$.
Furthermore, the lattice model is integrable since $\check R(u,q)$
satisfies the Yang-Baxter equation
\eqn\yangbaxter{(\check R(u;q)\otimes I )(I \otimes \check
R(u+v;q))(\check
R(v;q) \otimes I) \ = \ (I \otimes \check R(v;q) )(\check R(u+v;q)
\otimes
I) ( I \otimes  \check R(u;q))  \ . }
where $I$ is the identity matrix on one copy of $V$.

By going to the extreme anisotropic limit $(u \rightarrow 0)$, one
can
extract a ``spin-chain'' Hamiltonian from $\cal T$.
This is defined by
\eqn\hamilt{ {\cal H} \ = \ {\cal T}^{-1}(u) {\del \over \del u }
{\cal
T}(u) |_{u=0}  \ . }
An important property of $\cal H$ is that it has a non-trivial
boundary-term:
\eqn\boundary{ {\cal H}_{bdry} \ = \ - {2\over g} ~  \rho_G
\cdot ( h_1 - h_{2L}) \ , }
where $h_j$ is the Cartan sub-algebra
(C.S.A.) generator of $U_q(G)$ acting on the $j^{th}$
edge of the lattice.  This boundary term is essential for
$\cal H$ to commute with the quantum group \PS.
Moreover, this term has the effect of shifting all the ground state
energies exactly parallel to introducing a boundary charge
proportional to
$\rho_G$ into a gaussian model.

There are two equivalent height model or $IRF$ descriptions of the
foregoing vertex model, and both these descriptions can be viewed
as arising from a change of basis in ${\cal V}$
 \ref\pas{V.~ Pasquier, \cmp{118}
(1988) 365;  A.N.~ Kirillov and N.Yu.~Reshetikhin, in {\it
New Developments in the Theory of Knots}, editors: T.~Kohno and
N.J.~Teaneck, World Scientific (1989) (Advanced Series in
Mathematical Physics, v. 11).}\ref\AOc{ A.~ Ocneanu,
{\it Lond. Math. Soc. Lecture Notes Series} {\bf 136}.}.
The first height model, sometimes called the unrestricted height
model (or BCSOS model), is a completely trivial basis change:
the heights take values
on the entire weight lattice of $G$ and are assigned to vertices on
the lattice so that
the difference in height between two adjacent vertices is precisely
the weight of $V$ that is assigned to the interconnecting edge in
the vertex description \PA.  The second height description is much
better adapted to the quantum group.  One breaks the lattice into
time slices, each of which consists of a vertical zig-zag
of vertices and edges (see figure 1).   One starts at the top of
the lattice with some fixed representation of $G$ assigned to the
first  vertex in each zig-zag.
Representations, $W_p$, of $G$ are then associated to each vertex,
$p$, down a zig-zag so that $W_{p+1} \subseteq V \otimes W_p$.
One also keeps track of the total C.S.A. eigenvalue
(weight) of the state in the Hilbert space $\cal V$.  A list
of highest weights of representations $W_p$ and the total C.S.A.
eigenvalue along a zig-zag
determines a unique state in ${\cal V}$.  The set of
such states forms a basis for ${\cal V}$.
One thus has an association of representations of G to vertices on
the
lattice, and the $IRF$ Boltzmann weights can be
obtained by performing this basis
change on the transfer matrix \tranmatrix.  The fact that the
transfer
matrix of the vertex model commutes with $U_q(G)$ means that the
Boltzmann weights only care about the highest weight labelling of the
vertices and are independent of the overall C.S.A. charge.

Let  $k\in \ZZ$, $k\ge0$, and
suppose that
\eqn\rootcon{ q^{k+g+1} \ = \ \pm 1 }
where   $k+g+1$ is the smallest such integer.
One can now perform a quantum group truncation that leads to the
lattice
analogue of the $\Gkone$ models.
In the vertex model formulation the partition functions and
the correlation functions of the truncated
model can easily be written down in terms of the untruncated model.
One simply uses the modified trace \ref\Bax{R.J. ~Baxter,
S.B.~Kelland
and F.Y. ~Wu, {\it J. Phys.} {\bf A9} (1976) 397.}
\ref\VFRJ{V.F.R. ~Jones, {\it Bull. Am. Math. Soc. } {\bf 12}
(1985) 103.} \AOc\PA\PS
\eqn\truncation{Tr [{\cal O}] \ = \ tr [ {\cal O} \
\mu \otimes \cdots \otimes \mu ] \ ,} %
where
\eqn\defmuu{ \mu \ = \ q^{2 \rho_G \cdot h } \ . }
Once again, the insertion of the factor of $\mu $ can be thought of
as a
modification of the charge at infinity in the conformal model.

If one converts the foregoing into the height description that uses
the
representations $W_p$,  one finds that because of the modified trace,
only those representations with non-vanishing $q$-dimension
contribute to
partition and correlation functions.  Therefore one can
simply restrict the heights to those that correspond to the type
$II$ representations of $U_q(G)$ \PA\PS\SaZu .
Equivalently one can restrict to those weights that
are highest weights of the affine $G$ at level $k+1$.
The transfer matrix, of course, preserves such a truncation.  For
obvious
reasons this is called the restricted height or $IRF$ model.

One of the easiest methods of getting at the conformal field
theoretic limit of these lattice models is to consider the
unrestricted height model whose heights lie on the weight
lattice of $G$ \PA.  It can be argued that this model renormalizes
to a continuum limit consisting of $\ell$ free bosons (where $\ell$
is the rank of $G$) \FSZO\FSZ.
 The modified trace of the truncated
model, along  with appropriately chosen  spatial boundary terms,
can be argued to give the winding modes, or solitons of the gaussian
model certain phases that depend upon the winding numbers \IK.
In this way, one can arrive at the Poisson re-summed version of the
partition function \gaussiang\ with the correct phase factors \IK.
One can also reverse this procedure and try to use
the weight factors of a non-trivial modular invariant,
gaussian partition functions in order to
 determine the corresponding modifications to the lattice
theory.

To summarize,  one can think of the vertex formulation
(or the equivalent description in terms of unrestricted heights)
as being the direct counterpart of the continuum free bosonic theory.
The quantum group truncation is the counterpart of the
Feigen-Fuchs-Felder screening prescriptions.  The restricted height
model then emerges as the lattice counterpart of the
conformal model $\Gkone$, and it requires no further
modifications or truncation.  There are some important
subtleties about spatial boundary conditions, and we will comment
upon these at the end of this section.  Our strategy will therefore
be the following:  we will use the vertex/unrestricted height
description and its relation to the free bosonic theory to determine
the how to modify the $\check R$ matrix and how to further
modify the traces \truncation.  We will then change the basis
to the height model and pass to the restricted height model
and thus obtain the Boltzmann weights of a lattice model whose
continuum limit is \cosetmodel.

The first step to getting the lattice analogue of the model
\cosetmodel\ is to take the vertex desciption of the $\Gkone$
models, set $k=0$ and untwist the transfer matrix.  This can be
done by ``conjugating'' \foot{
Note, this is not really a true conjugation
of the $\check R$-matrix as the left hand factor in (3.11)
is not the inverse of the right hand factor in (3.11).}
the $\check R$-matrix of
$G$ \ref\BL{D.~Bernard and A.~Le~Clair, Commun. Math. Phys.{\bf 142}
(1991)
91.} \ref\LNW{A.~Le~Clair, D.~Nemeschansky and N.P.~Warner, in
preparation.}.
Define:
\eqn\rcheckprime{ \check R^\prime(x,q) \ = \ \Bigl [ 1 \otimes x^{-
{2
\over g }( \rho_G-\rho_H) \cdot  h} \Bigr ] \check R(x,q)
\Bigl
[ x^{{2\over g } ( \rho_G- \rho_H) \cdot h} \otimes 1 \Bigr ] \ . }
This $\check R^\prime$-matrix has several important properties.
It, of course, still satisfies the Yang-Baxter equations \yangbaxter
{}.

It commutes with $U_q(H^\prime)$, where  $H^\prime$
is the semi-simple factor of $H$ ({\it i.e.}
$H = H^\prime \times U(1)$), because $(\rho_G-\rho_H)
\cdot  h$ defines the $U(1)$ that is orthogonal to $H^\prime$.
If one employs $\check R^\prime$  to construct the transfer matrix,
one can easily verify that the net effect of ``conjugation'' is
to simply add boundary terms to  the transfer matrix ${\cal T} (u)$.
In particular, the analogue of \hamilt\
has a boundary term of the form \boundary , but with
$\rho_G$ replaced by $\rho_H$.
This yields precisely the required shift in the ground state energy
to go
from the central charge, $c=0$, of the topological
${\cal G}_{0,1}$ theory
to the correct value of $c$ for the $N=2$ supersymmetric
coset model.

The only other step that is required to obtain the lattice analogue
of
\cosetmodel\ is to re-examine the modified trace \truncation\ in the
light of our earlier comments.  From the comparison of the $N=2$
superconformal partition function with that
 of the $\Gkone$ model, we see that the untwisted
${\cal G}_{0,1}$ partition function gives the $N=2$ superconformal
partition function in the Ramond sector with an insertion of
$(-1)^F$.
In the sum over solitons in \poissonr\ the operator $(-1)^F$
corresponds to weighting a soliton by a phase:
\eqn\phase{e^{{2\pi i \over \sqrt{(g+1) g} }(\rho_G-\rho_H)\cdot x }
\ , }
where $x$ is the winding vector in the timelike direction.
Therefore,
to obtain lattice partition functions and correlation functions
without
such an insertion of $(-1)^F$, we must remove this phase from the
modified trace  of \truncation.  On the lattice Hilbert space, ${\cal
V}$,
the operator $(-1)^F$ is simply:
\eqn\onetof{\Delta ( q^{-2 (\rho_G-\rho_H)\cdot h } ) \ , }
where $\Delta$ is the co-product
\foot{Given the form of the phase $(-1)^F$,
this operator should correspond to something of the form
$\Delta(q^{a(\rho_G-\rho_H)\cdot h})$, where a is a  constant.
It will soon become evident as to why one has $a=-2$.}.
Making a further insertion of \onetof\ into
\truncation\ merely amounts to replacing the modified trace
\truncation\ by the $H$-modified trace:
\eqn\modifiedtr{ Tr_H({\cal O }) \ = \ tr [{\cal O } \ \mu_H
\otimes \cdots \otimes \mu_H] \ , }
where
\eqn\muhdef{\mu_H \ = \ q^{2\rho_H \cdot  h} \ . }
If one converts this to the height formulation, this new $H$-modified
trace means that the only those heights that correspond to
type $II$ representations of $U_q(H^\prime)$ will contribute.

Thus the essential idea is that we are
using a vertex model based upon $G$, but only performing  a quantum
group
truncation with respect to $H^\prime$.
Since one has $q=e^{i {\pi \over g+1}}$ this means that one
truncates $H^\prime$ highest weights to those weights that
are highest weight labels of affine $H^\prime$ at level $g-h+1$.
The states that lie purely in the Hilbert space of the $U(1)$ factor
of $H$
are completely unaffected by the truncation and the boundary charge.
This direction still corresponds to an unrestricted, free $U(1)$.

Perhaps the most convincing argument as to why the foregoing
construction
is the correct one is obtained by reconsidering, and expanding upon,
our
discussion of the conformal field theory defined
by \cosetmodel.  Because the
factor of $G$ in \cosetmodel\ is of level one, it can be replaced
by a factor of $H$ at level one (because the rank of $G$ is equal to
the
rank of $H$).  In addition, the fact that the embedding of $H$ into
$SO_1(dim(G/H))$ is conformal means that one can replace the factor
of $SO(dim(G/H))$ by $H_{g-h}$.  The important proviso is that one
can make these replacements provided that one restricts the
representations of $H_1$ and $H_{g-h}$ to those combinations that
make up the representations of $G_1$ and $SO_1(dim(G/H))$.  Hence,
modulo
this statement about represenations, one is dealing with a coset:
$H_1 \times H_{g-h}/H_{g-h+1}$.  Now recall that $H = H^\prime
\times U(1)$, where $H^\prime$ is semi-simple, and once again, modulo
the careful treatment of the radii of $U(1)$ factors and the correct
association of $U(1)$ charges with representations of $H^\prime$,
one can cancel the $U(1)$ factors between the numerator and
denominator
and see that one is really working with a conformal coset model:
\eqn\hcoset{ {H^\prime_1 \times H^\prime_{g-h} \over H^\prime_{g-h+1}
}
\times U(1) \ . }

The $U(1)$ factor in \hcoset\ is precisely the $N=2$, $U(1)$ current
\uonecurrent.  Much of the labour in section 2 was spent upon
deriving exactly which Hilbert spaces of the $H^\prime$-coset
model \hcoset\ were to be employed, and determining the associated
$N=2$, $U(1)$ charges so that one would obtain \cosetmodel.  The
result of this labour may be summarized as follows.
One uses \deflambda\ in $H_{g-h}$ for the Ramond sector,
while one uses
\eqn\nslabels{\lambda(\alpha,w) \ = \ w(\rho_G) -\rho_G +g \alpha }
in $H_{g-h}$ in the Neveu-Schwarz sector.  In the $H_1$ factor one
simply employs the weights of $H$ that are, in fact, weights of $G$.
Now observe that vectors in \nslabels\ are actually roots of $G$,
and that the vectors in \deflambda\ are roots of $G$ shifted by
$\rho_G - \rho_H$.  Consequently one can roughly think of
\cosetmodel\
as \hcoset\ with a selection rule on the $U(1)$ charge that
amounts to requiring that the $U(1)$ charges are added so that the
weights of $H^\prime$ extend to weights of $G$, or at least weights
of $G$ up to a possible shift by $\rho_G - \rho_H$.  This selection
rule is also evident in the partition function \poissonr, where the
summation is over winding modes on $\binv M(G)$.

The new vertex model introduced above has been constructed in such a
manner that it also exhibits the foregoing
$H^\prime$-coset structure.  First, the corresponding
unrestricted height model has a continuum limit that
is described by $\ell$ free bosons.  We have ensured that the
transfer
matrix commutes with $U_q(H^\prime)$, and that the spin-chain
hamiltonian only has non-hermitian boundary terms in the
$H^\prime$ direction.   We have arranged the modified trace so as
to only perform the $H^\prime$ quantum group truncation, and the
value
$q$:
\eqn\qvalue{ q ~\equiv ~ e^{{i \pi }\over {g+1}}}
is precisely the correct one to obtain \hcoset\ at the right level.
As remarked above, the states that lie in the
Hilbert space of the $U(1)$ factor are completely unaffected
by the truncation and boundary charge, and this
direction thus corresponds to an unrestricted, free $U(1)$.
The fact that we have built the model starting with the $\check R$
matrices of $G$, and heights that are  weights of $G$ means that
the continuum limit will have $U(1)$ charges associated with
$H^\prime$
representations in accordance with the selection rule described
above.
These facts all give us considerable confidence that we have indeed
given a lattice description of a model whose
continuum limit is \cosetmodel.

It is now relatively simple to summarize the restricted
height formulation of the $N=2$ supersymmetric models.
The $IRF$ formulation of the $\Gkone$ models can be
described in terms of a graph generated by the fusion rules  of
$G$ at level $k$.
In particular, one generates a directed graph by considering which
representation can be connected by fusion with the fundamental
representation, $V$ \FZU .
For the $N=2$ models  the situation is very similar, except,
it works on the graph defined by all the weights of $G$ that are
highest weights of
$H^\prime$ at level $g-h+1$.
(The graph may be infinite in one direction because the $U(1)$ charge
is
arbitrary).
One can make it a directed graph by considering which vertices can be
connected by fusions with $V$.
It should, of course be remembered that $V$ will be decomposable into
at
least two $H$ representations, and one must consider all the
irreducible
pieces in describing the  fusion graph.
The Boltzmann weights in the height description can be computed
using the basis change that underlies the
$IRF \leftrightarrow$ vertex correspondence, but almost all of the
hard
work can be circumvented by using the known solution for the $\Gkone$
models.

As an example, consider the grassmannian models, which are
 of the form \cosetmodel\
with $G=SU(m+n)$ and $H=SU(m) \times SU(n) \times U(1)$.
The $\check R$ matrix is given by \chekrmatic, but with
$\alpha, \beta$ running from $1$ to $m+n$.  Let  the indices
$a,b$ and $i,j$ run from $1$ to $m$ and
from $m+1$ to $m+n$ respectively. Let $\check R_{(1)}(u,q)$ and
$\check R_{(2)}(u,q)$ be the diagonal $m \times m$ and
$n \times n$ blocks in $\check R$.
Note that the sub-matrices $\check R_{(1)}$ and $\check R_{(2)}$
are simply the $\check R$-matrices for $SU(m)$ and $SU(n)$
respectively.  Under the ``conjugation'' operation
\rcheckprime, these sub-matrices are not modified.  One can also
easily
verify that the only part of \chekrmatic\ that is modified is the
third
term for $\alpha > m, \beta \le m$ or $\alpha \le m, \beta > m$.
Indeed, one finds that in $\check R^\prime$, with $\alpha$ and
$\beta$
in the foregoing index ranges, this third term reduces to:
$$
(q-q^{-1} )\Bigl [ \sum_{ \alpha > m, \beta \le m}  +
\sum_{\alpha \le m, \beta > m} \Bigr ] E_{\alpha \alpha }
\otimes E_{\beta\beta } \ ,
$$
It follows immediately that
\eqn\offdiaga{\eqalign{\check R^\prime(u,q)_{a i, b j} &~=~
\check R^\prime(u,q)_{i a, j b} ~=~ (q ~-~ q^{-1}) ~\delta_{ab}
{}~\delta_{ij} \cr  \check R^\prime(u,q)_{i a, b j} &~=~
\check R^\prime(u,q)_{a i, j b} ~=~ (x ~-~ x^{-1}) ~\delta_{ab}
{}~\delta_{ij}. \cr}}
Converting the face transfer matrix \defxi\ into the $IRF$ language
is now
elementary.
Consider the face with assigned heights $( \Lambda_{p-1},
\Lambda_p, \Lambda_{p+1},  \Lambda^{\prime}_p)$, as shown in figure
2.
Decompose each height according to $\Lambda \equiv (\lambda, \nu
;q)$, where
$\lambda$ is a highest weight of $SU_n(m)$, $\nu$ is a highest weight
of
$SU_m(n)$, and $q$ is the $U(1)$ charge: $q= 2 (\rho_G-\rho_H) \cdot
\Lambda$.
Because of the foregoing block decomposition of $\check R^\prime$,
the height model transfer matrix has three ways in which it can act:
\hfill\break
\noindent
(i) The $\lambda$ label can evolve exactly as it does in the
$SU_1(m)\times
SU_n(m) /SU_{n+1}(m)$ height model, with $q_p^{\prime} = q_p$ and
$\nu^{\prime}
_p = \nu_p $.
\hfill\break
\noindent
(ii) The $\nu$ label can evolve exactly as it does in
the $SU_1(n) \times SU_m(n) /SU_{m+1}(n)$ height model, with
$q_p^{\prime}
= q_p $ and $\lambda_p^{\prime} = \lambda_p $.
\hfill\break
\noindent
(iii) The off-diagonal terms of \offdiaga\ act.

\epsfxsize = 3.4in
\vbox{\vskip -.3in \hbox{\centerline{\epsffile{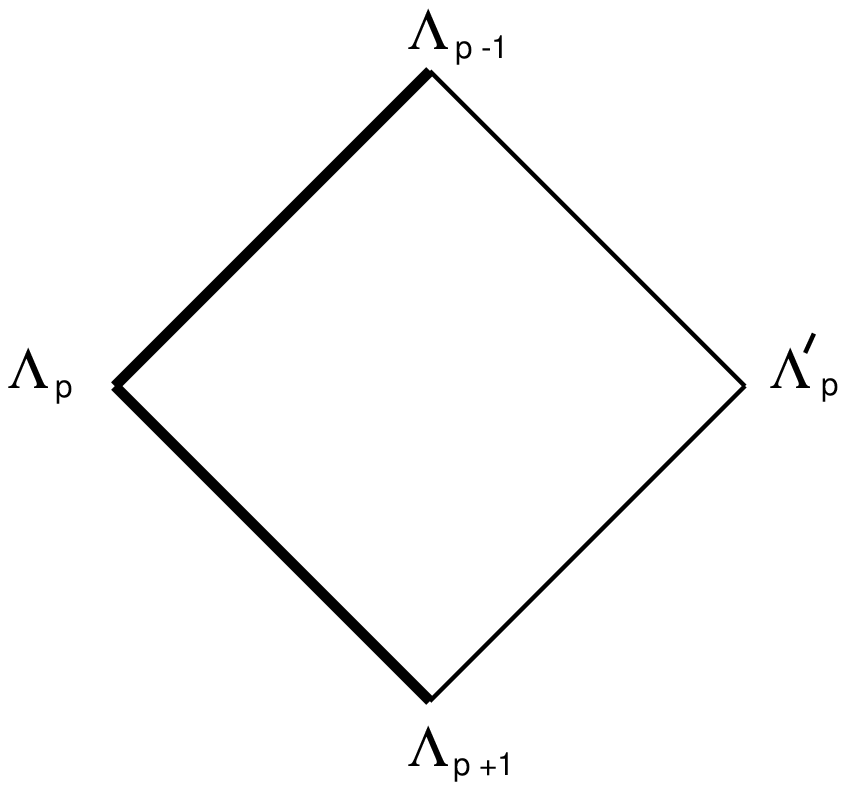}}}
{\vskip -.4in \leftskip .5in \rightskip .5in \noindent \ninerm
\baselineskip=10pt Figure 2.
In the $IRF$ formulation the heights, $\Lambda$, are
associated to vertices as shown. \smallskip}}

The expression for $X_p(u,q)$ in situations $(i)$ and $(ii)$ is well
known
(see, for example, \ref\HeW{H.~Wenzel, Representations of Hecke
Algebras and
Subfactors, Univ. of Pennsylvania thesis (1985);
{\it Invent. Math.} {\bf 92} (1988) 349. }\FZU ).
One can write
\eqn\defxpu{X_p(u,q) \ = \ sin(\gamma + u ) I_p - sin(u) U_p \  , }
where $q=e^{i \gamma}$ and $x=e^{iu}$.
Introduce the vectors $e_1 \equiv \nu_1$, $e_n \equiv  - \nu_{n-1}$
and $e_j \equiv \nu_j - \nu_{j-1}$  for $ j=2, \ldots , n-1$,
where $\nu_j$ is the $j^{th}$ fundamental weight of  $SU(n)$,
and define a function
\eqn\defsjk{ s_{jk}(\nu) \ \equiv  \ sin\Bigl ( {\pi \over m+n+1}
(e_j-e_k) \cdot \nu \Bigr )  \ . }
The operator, $U_p$, then has the form:
\eqn\defup{ U_p \ \equiv \ (1-\delta_{jl}) { \Bigl (s_{jl}(\nu+e_j)
s_{jl}(\nu+e_k)\Bigr )^{\half} \over s_{jl}(\nu) } \ ,  }
where $\Lambda_{p-1} \equiv (\lambda, \nu ; q )$,
$\Lambda_p \equiv (\lambda, \nu+e_j
; q-m)$, $\Lambda_{p+1} \equiv (\lambda, \nu+e_j+e_l; q-2m)$ and
$\Lambda^{\prime}_p \equiv (\lambda , \nu +e_k ;q-m)$.
The evolution in the $SU(m)$ factor is similar.
One should also note that if $n=1$, the evolution in the ``$SU(1)$''
direction
merely involves a shift of the $U(1)$ charge and, as can
be seen from  \chekrmatic, the transfer matrix
is simply a multiplicative factor  of $sin(\gamma+u )$
({\it i.e.} the matrix $U_p$ is  zero ).

In the foregoing components of the transfer matrix one either had
$\lambda_{p-1}
= \lambda_p = \lambda_{p+1} $ or $\nu_{p-1} = \nu_p = \nu_{p+1} $.
The off-diagonal parts of $\check R^{\prime}$ deal with the evolution
when
neither of these equalities hold.
However, for $X_p(u,q)$ to be non-zero, one must  still have either
(i) $ \lambda_p = \lambda_{p-1}$ and $ \nu_{p+1} = \nu_p $ or
(ii) $\nu_p = \nu_{p-1} $ and $\lambda_{p+1} = \lambda_p $.
In which case one has
\eqn\xpe{ X_p(u,q)  \ = \ sin (\gamma) \  I + sin(u) \  E \ , }
where $E$ is an operator that is equal to 1 if
(i) $\lambda^{\prime}_p = \lambda_{p+1} $ and $ \nu^{\prime}_p =
\nu_{p-1}$,
or (ii) $ \lambda^{\prime}_p = \lambda_{p-1}$ and $\nu^{\prime}_p =
\nu_{p+1}$,
and $E$ vanishes otherwise.
Putting it more directly:
If $\Lambda_{p-1} = (\lambda_{p-1} , \nu_{p-1}; q)$,
$\Lambda_p = (\lambda_{p-1}, \nu_p ;q-m)$ and
$\Lambda_{p+1} = (\lambda_{p+1}, \nu_p ; q+n-m) $
then $X_p(u,q)$ is $sin(\gamma)$ or $sin(u)  $ depending upon whether
$\Lambda^\prime_p = (\lambda_{p-1}, \nu_p ;q-m)$ or
$\Lambda_p^{\prime} =
(\lambda_{p+1}, \nu_{p-1}, q+n)$ respectively.
If $\Lambda_{p-1} = (\lambda_{p-1}, \nu_{p-1}; q)$, $\Lambda_p =
(\lambda_p,
\nu_{p-1} ;q+n)$ and $ \Lambda_{p+1} = ( \lambda_p , \nu_{p+1};
q+n-m)$
then $X_p(u;q)$ is $sin (\gamma)$ or $ sin (u)$ depending upon
whether
$\Lambda^{\prime}_p \equiv ( \lambda_p, \nu_{p-1} ; q+n)$ or
$\Lambda^{\prime}_p \equiv (\lambda_{p-1}, \nu_{p+1} ;q-m)$
respectively.

Before concluding this section we think it important to make some
remarks
about spatial boundary conditions.  In the lattice analogues of the
$\Gkone$ models the relation between lattice boundary conditions
and sectors of the continuum limit is rather subtle.
One should begin by noting that the vertex model must necessarily
have
free boundary conditions if it is to commute with $U_q(G)$ \PS.
If one now constructs the partition function by using the modified
trace \truncation\ and taking $Z_v = Tr({\cal T}^n)$, then the
spatial
and temporal boundary conditions are very different, and $Z_v$ will
not be modular invariant in the continuum limit.  It is of obvious
interest to determine which combination of characters one gets from
$Z_v$ in the continuum limit.  A detailed analysis for $SU(2)$ can
be found in \PS.  The results suggest that in general $Z_v$ will
be a particular combination of characters $\chi_{0,\Lambda}$, where
the subscripts connote the affine labels of $G_k$ and $G_{k+1}$
in the $\Gkone$ model.  Intuitively one can understand this as
follows.
The choice of the value of $q$ in \rootcon\ suggests that the lattice
quantum group should be identified with the screening charges
obtained
from the operators $e^{i \beta \alpha_j \cdot \phi}$, as opposed to
the operators $e^{i \binv \alpha_j \cdot \phi}$.  This is because the
former operators have braiding relationships that involve the
$(k+g+1)^{th}$ roots of unity whereas the latter have braiding
relationships that involve the $(k+g)^{th}$ roots of unity.  The
physical states on the lattice are constructed from  non-trivial
representations of the screening algebra and so one should expect
the same thing in the continuum limit.  This suggests that the
Hilbert spaces should be those built from the primary field
$\Phi_{0,\Lambda}$ in \gprims.  Putting it
somewhat differently, if one considers the lattice states that
remain after the quantum group truncation, the corresponding
heights are precisely the affine highest weights of $G_{k+1}$
(and not $G_k$).

While $Z_v$ is a very particular combination of
the abovementionned characters, one can also extract a particular
one of these characters by restricting the modified trace
to those states for which the last representation, $W_{2L}$, in
the sequence described earlier is fixed
to a given representation (with
highest weight corresponding to an affine highest weights of
$G_{k+1}$) \PS.  One can also go beyond the restricted class
of characters by imposing spatially periodic boundary
conditions and making the appropriate modifications to
the transfer matrix.  The partition function then gets
contributions from all sectors of the theory, but one loses
the simple quantum group invariance of the transfer matrix.  With
a considerable amount of hard work, one can recover a more
exotic form of the quantum group structure and use it to see
how each sector is accounted for in the total partition function \PS.

There are parallel situations in the $IRF$ descriptions of these
models.   For example, if one fixes the heights on the top and
bottom of the lattice then, in the continuum limit, one gets
partition functions that enumerate the subclass of characters,
$\chi_{0,\Lambda}$, mentionned above.
On the other hand, if one merely wishes to
construct a modular invariant partition function for the
lattice model of interest it is elementary to accomplish
this in the restricted height formulation.  One does the
obvious thing and uses the Boltzmann
weights of the model on a toroidal lattice ({\it i.e.} one that
is periodically identified in space and time).  Since the
restricted height model needs no further modifications or
truncations, one need not introduce any factors or phases at
the boundaries.   In the continuum limit one will get the
diagonal modular invariant for $\Gkone$ or
$G \times SO(dim(G/H))/H$ depending upon the choice of the
restricted Boltzmann weights.  The purpose of taking the
circuitous route through vertex and unrestricted height
models is that this approach gives us a computational method
of deducing the Boltzmann weights in the resticted $IRF$
model.

As a final comment on the subject of boundary conditions,
we expect that the extraction of the separate sectors of
\cosetmodel\ will be no more complicated than it is for
the $\Gkone$ models.  As was observed in the previous section,
the only difference in the unrestricted height model is the
insertion of an extremely simple phases \newphase\ in the spatial
direction of the soliton sum.

\newsec{Lattice operators}

Having seen how closely related the $SLOHSS$ models are to the
lattice
models based upon $G$, it is easy to translate other lattice results
to the
$N=2$ supersymmetric theory.
Consider, for instance the order parameters \PA.
For the $\Gkone$ models, the natural order parameters
renormalize to vertex operators
\eqn\vertexprim{\psi_{\Lambda}(z,\bar z) \ = \ e^{ i \alpha_0 \Lambda
\cdot
\phi(z,\bar z) } \ , }
where
\eqn\defalpha{ \alpha_0 \ = \ { 1 \over \sqrt {(k+g)(k+g+1) } }  }
is the charge at infinity, and $\Lambda$ is the highest weight label
of
$G$ at level $k+1$.  Since the $N=2$ supersymmetric model
can be written in the form \hcoset, the
order parameters of the lattice model will renormalize
to a similar vertex operator to \vertexprim, but with $\Lambda$
constructed out of the correct combination of $H^\prime$ weights and
$U(1)$ charges.  Now recall that in the Neveu-Schwarz
sector, the $H$ labels that come from $SO(dim(G/H))$ are
given by \nslabels. One should recall that in equation \nslabels\
one has $w\in W(G)/W(H)$ and $ \alpha \in M(G) /M(H) $.
For operators of the form \vertexprim,
the translation by $\alpha$ can be neglected since it represents a
a trivial automorphism generated by spectral flow \LVW.
It follows that the set of natural order parameters of the
lattice model will renormalize to
\eqn\orderpara{e^{ {i \over \sqrt{g(g+1) } }(w(\rho_G) - \rho_G)
\cdot \phi(z) } \ .}
As was noted in section 2, these are precisely  the chiral
primary fields of the $N=2$ theory.

There are also further operators of interest that arise directly
from the quantum group structure.  There are two generators,
$X_{\pm \gamma}$, of the quantum group $G$ that are not generators
of quantum $H^\prime$.  These generators obey the commutator
\eqn\quancom{ [X_{\gamma} , X_{-\gamma} ] \ = \ { q^{\gamma \cdot h
}-
q^{-\gamma \cdot h} \over q- q^{-1} } \ , }
where $\gamma$ is the simple root of $G$ that extends the simple root
system of $H^\prime$ to one for $G$.
One should recall that the original $\check R$ matrix not only
commutes with $\Delta(X_{\pm \alpha_j})$
for all simple roots $\alpha_j$ of $G$, but also satisfies
\eqn\affqgp{ \eqalign{\check R  \ &\big( X_{\pm \psi} \otimes
q^{+\half \psi \cdot h} ~+~ q^{-\half \psi \cdot h} \otimes
\big( x^{\pm 2}~ X_{\pm \psi}\big) \big)\cr  ~=~ &\big(
\big(x^{\pm 2} ~ X_{\pm \psi} \big)
\otimes q^{+\half \psi \cdot h} ~+~ q^{-\half \psi \cdot h} \otimes
 X_{\pm \psi} \big) ~\check R \ . \cr}}
This means that the matrix $\check R^\prime$ satisfies a similar
equation:
\eqn\hssqgp{ \eqalign{\check R^\prime \ &\big( X_{\pm \alpha} \otimes
q^{-\half \alpha \cdot h} ~+~ q^{+\half \alpha \cdot h} \otimes
\big( x^{\pm 1}~ X_{\pm \alpha}\big) \big)\cr  ~=~ &\big(
\big(x^{\pm 1} ~ X_{\pm \alpha} \big)
\otimes q^{-\half \alpha \cdot h} ~+~ q^{+\half \alpha \cdot h}
\otimes
 X_{\pm \alpha} \big) ~\check R^\prime \ . \cr}}
for both $\alpha = \gamma$ and for $\alpha = - \psi$.
Given the close relationship between the quantum group
generators of the lattice model and the screening charges
of the continuum theory, one would expect that these four
affine quantum group generators, $X_{\pm \gamma}$ and
$X_{\pm \psi}$, should, in the contiuum limit, be identified
with a subset of the vertex operators:
$e^{i \beta \gamma \cdot \phi(z)}$, $e^{-i \binv
\gamma \cdot \phi (z)}$, $e^{-i \beta \psi \cdot \phi(z)}$,
$e^{i \binv \psi \cdot \phi (z)}$, and their anti-holomorphic
counterparts.  To make the proper identification one should
note that $X_{\gamma}$ and $X_{\psi}$ have the same $N=2$,
$U(1)$ charge and this charge is equal and opposite to that
of $X_{-\gamma}$ and $X_{-\psi}$.  This means that the
continuum vertex operators should either all involve the
coupling constant $\beta$ or should all involve $\binv$.  The
fact that $q^{k+g+1} = 1$ in the lattice model suggests that
we should look for the same root of unity in braiding relations
of the vertex operators.   This then leads us to relate
$X_{\gamma}$ and $X_{-\gamma}$ to
$e^{i \beta \gamma \cdot \phi(z)}$ and $e^{i \beta
\gamma \cdot \phi (\bar z)}$ respectively, and
$X_{-\psi}$ and $X_{\psi}$ to
$e^{-i \beta \psi \cdot \phi(z)}$ and
$e^{- i \beta \psi \cdot \phi (\bar z)}$ respectively
\foot{One  should remember that for supersymmetric
perturbations of the conformal field theory, the non-trivial,
off-critical conserved $U(1)$ charge is $Q= J_0 - \tilde J_0$.
This means that holomorphic and anti-holomorphic operators
have an extra relative sign for their $U(1)$ charges. }.
In the continuum field theory, these four operators
correspond to $(\Gminus \Phi)(z)$, $(\widetilde \Gminus \Phi)(\bar
z)$,
$(\Gplus \bar \Phi)(z)$ and $(\widetilde \Gplus \bar \Phi )(\bar z)$,
where $\Phi$ is the most relevant chiral primary field and $\bar \Phi
$
is its anti-chiral conjugate.

The foregoing operators have extremely special properties.  When
used to perturb the conformal model, they yield massive, integrable
field theories \FMVW\FLMW\NW.
These operators also generate the perturbation that is needed to take
the topological, twisted $N=2$ superconformal model
to the toplogical $G_0 \times G_1/G_1$ model in such a way that
the toplogical correlation functions of \ref\DVV{R.~Dijkgraaf,
E.~Verlinde and H.~Verlinde, \nup{352} (1991) 59.} yield the fusion
rules.  It is interesting to note that the foregoing operators
also become a component of the supercurrent in these massive
off-critical models.

The fact that the lattice analogues of the operators that give rise
to
integrable field theories are precisely the operators that extend
$U_q(H^\prime)$ to the affine quantum group $\widehat {U_q(G)}$
suggests that the same might be true in the continuum.  That is, if
one
uses a free field description of the conformal model, then the
screening prescription should combine with the conformal
perturbation theory to produce a realization of $\widehat {U_q(G)}$.
This phenomenon has already been observed in the integrable
field theories that appear as perturbations of the non-supersymmetric
minimal models \ref\SMath{S.~Mathur, \nup{369} (1992) 433.}.
For the vertex model analogues of these non-supersymmetric
models, the generators of $\widehat {U_q(SU(2))}$ can be  directly
related to the vertex operators of a screening current and of the
integrable pertubation.
Consequently, the lattice affine quantum group gives us some
further understanding of the results of \SMath, and
also leads to an intriguing prediction for the structure
of the conformal  perturbation expansion in the perturbed
$SLOHSS$ models.

\newsec{Conclusions}

It is evident that one can find simple lattice formulations of $N=2$
supersymmetric $SLOHSS$ models by making a straightforward
modification of
the $\Gkone$ models.
We also find it satisfying that the order parameters of the
model become the chiral primary fields become the \LG fields of the
$N=2$ superconformal model in the continuum limit. The role played
by the operators that extend $U_q(H^\prime)$ to $\widehat {U_q(G)}$
also provides a new perspective on the continuum integrable field
theories.

The technique that we have used here is basically
to formulate  a model using a
group $G$, and then quantum group truncate with respect to a subgroup
$H$.
This procedure obviously admits generalizations
(see, for example \ref\MW{S.~Mathur and N.~P.~Warner
\plt{254} (1991) 365. }).
As in \MW\ the resulting theory will probably only be unitary when
$G/H$ is a
symmetric space \foot{ Note that $G/H$ does not need to be
a {\it hermitian} symmetric space in order to produce a unitary
theory.
A non-hermitian symmetric space yields an $N=1$ supersymmetric
theory.}.
On the other hand, lack of unitarity has never seemed to be an
impediment
to finding physically interesting statistical mechanics models, and
so
there may  well be some interesting non-unitary, as well as unitary,
generalizations.

There are also several other clear directions for further research.
It would be valuable to perform some Bethe Ansatz calculations to
confirm
that the lattice models described  here do indeed yield the correct
conformal
weights for primary fields.  It is also very probable
that the fused vertex models, and their $IRF$ counterparts,
can be modified to obtain a formulation of the general $N=2$
supersymmetric coset models
\eqn\gencoset{{ G_k \times SO(dim(G/H)) \over H} \ . }
It would be interesting to examine the details of how this works.
In this paper we have also only examined the critical model,
with Boltzmann weights tuned to the critical temperature.
We next plan to obtain the off-critical Boltzmann weights for the
lattice analogues of the $N=2$ supersymmetric models.
Among our aims in doing this is to use the corner transfer matrix
methods
to understand the critical model more precisely.
We are also extremely curious to see whether we can find the lattice
antecedents of the supersymmetry generators.

Finally, there is the question of the lattice analogues of
the topological matter models that can be obtained from the
coset models \cosetmodel, and more generally, \gencoset.
One should recall that the first step in our construction of the
new vertex models was to take the vertex model constructed
from the $\check R$ matrix of $G$, and set
$q = e^{{i \pi} \over {g+1}}$ ({\it i.e. } put $k=0$).  The
$U_q(G)$ quantum group structure is supposedly trivial for this
value of $q$. It is however one of the lessons of the topological
field theory that trivial representations often combine to
make physically interesting theories. (Or to paraphrase
Stanislaw Lem ``Everybody knows that non-trivial representations
do not exist, but each one does it in a different way.'')
One finds that with this value of $q$, the only $G$ reperesentations
with non-vanishing $q$-dimension are those that
correspond to representations of  affine $G$ at of level one.
When one passes to the
restricted height model, the directed fusion graph \FZU\ is
trivial since each representation of $G_1$ fuses with $V$
to yield an unique result.  This means that all the lattice
heights are fixed once one has chosen one of them.  This virtually
trivial lattice model is what we will refer to as the topological
lattice model because, in the continuum limit, it is precisely the
analogue of the perturbed, topologically twisted $N=2$ superconformal
theory that was discussed in \NW.  For the $\ell$-fused vertex model,
or its $IRF$ equivalent,  the topological model has
$q = e^{{i \pi} \over {\ell + g}}$ ({\it i.e. } $k=0$).
Once again, all the lattice heights are fixed by the choice of a
single
height somewhere on the lattice.  This is because the directed
fusion graph of the restricted height model is defined using a
particular level $\ell$ representation, $V$, of $G$, and one can
easily verify that the highest weight of $V$
defines a simple current of $G_\ell$ \ref\SYSC{A.N.~Schellekens
and S.~Yankielowicz, \nup{327} (1989) 673; \plt{227} (1989) 387.}
\ref\KInt{ K.~Intriligator, \nup{332} (1990) 541.}.  This means that
each representation of $G_\ell$ fuses with $V$ to generate an
unique result.  This rigid structure of the fusion graph
of the topological lattice model leads us to expect
that the $N$-point correlation functions
of such a model will all be constants, and
that the three point functions will
reproduce the fusion algebra of $G_\ell$.
In spite of the fact that this is intuitively
very reasonable, it is still necessary
to check the details, and perhaps more interesting, to see how such a
topological sector embeds in the physical model.

\medskip

\noindent
{\bf Acknowledgements}

\smallskip

We would like to thank H.~Saleur for extensive discussion and
education on the
subject of lattice models, and for his comments on an early version
of the manuscript.  We are also grateful to J.-B~Zuber for
communication
about height models and to  A.~Le Clair for discussions on
$R$-matrices and
fractional supercharges.

\vfill
\listrefs
\eject
\end